\documentclass[a4paper,prd,showpacs,noshowkeys,preprint]{revtex4}
\usepackage[latin1]{inputenc}
\usepackage{amsmath}
\usepackage{mathrsfs}

\providecommand{\bp}{{\bf p}}

\providecommand{\bq}{{\bf q}}
\providecommand{\bx}{{\bf x}}
\providecommand{\by}{{\bf y}}
\providecommand{\br}{{\bf r}}
\providecommand{\bR}{{\bf R}}
\providecommand{\TT}{\mathsf{T}}
\providecommand{\ST}{\mt{ST}}
\providecommand{\lag}{\mathscr{L}}
\providecommand{\sla}[1]{~\hs{-1.5ex}\not\hs{-.4ex}#1\hs{.1ex}}

\providecommand{\eq}[1]{\begin{equation} #1 \end{equation}}
\providecommand{\eqarr}[1]{\begin{eqnarray} #1 \end{eqnarray}}

\providecommand{\mt}[1]{\mbox{\tiny $#1$}}
\providecommand{\ms}[1]{\mbox{\small $#1$}}
\providecommand{\mn}[1]{\mbox{\normalsize $#1$}}
\providecommand{\mss}[1]{\mbox{\scriptsize $#1$}}
\providecommand{\mfn}[1]{\mbox{\footnotesize $#1$}}
\providecommand{\mcal}[1]{\mathcal{#1}}
\providecommand{\bs}[1]{\boldsymbol{#1}}
\providecommand{\hs}[1]{\hspace{#1}}
\providecommand{\ket}[1]{\vert #1 \rangle}
\providecommand{\bra}[1]{\langle #1 \vert}
\providecommand{\braket}[2]{\langle #1\vert #2 \rangle}
\providecommand{\aver}[1]{\langle #1 \rangle}
\def\pr{\mbox{\tiny $\prime$}}

\font\bb=bbmss12 scaled 1000
\def\id{\mbox{\bb 1}}

\begin{document}
\title{First quantized approaches to neutrino oscillations
and second quantization}
\author{C.~C.~Nishi}
\email{ccnishi@ift.unesp.br}
\affiliation{Instituto de F\'\i sica Te\'orica,
UNESP -- São Paulo State University
\\
Rua Pamplona, 145\\
01405-900 -- S\~ao Paulo, Brazil
}


\begin{abstract}
Neutrino oscillations are treated from the point of view of relativistic first
quantized theories and compared to second quantized treatments. Within first
quantized theories, general oscillation probabilities can be found for Dirac
fermions and charged spin 0 bosons. A clear modification in the oscillation
formulas can be obtained and its origin is elucidated and confirmed to be
inevitable from completeness and causality requirements. The left-handed nature
of created and detected neutrinos can also be implemented in the first quantized
Dirac theory in presence of mixing; the probability loss due to the changing of
initially left-handed neutrinos to the undetected right-handed neutrinos can be
obtained in analytic form.
Concerning second quantized approaches, it is shown in a calculation using
virtual neutrino propagation that both neutrinos and antineutrinos may also
contribute as intermediate particles. The sign of the contributing neutrino
energy may have to be chosen explicitly without being automatic in the
formalism. At last, a simple second quantized description of the flavor
oscillation phenomenon is devised. In this description there is no interference
terms between positive and negative components, but it still gives simple
normalized oscillation probabilities. A new effect appearing in this
context is an inevitable but tiny violation of the initial flavor of neutrinos.
The probability loss due to the conversion of left-handed neutrinos to
right-handed neutrinos is also presented.

\end{abstract}
\pacs{03.65.Pm, 14.60.Pq, 03.70.+k}
\keywords{Neutrino oscillations, relativistic quantum mechanics, quantum fields}
\maketitle
\section{Introduction}
\label{sec:intro}

Compelling experimental evidences~\cite{exp} have shown that
neutrinos undergo flavor oscillations in vacuum. Consequently, this fact
requires massive neutrinos with mixing. These ingredients are not present
in the standard model of elementary particles. For this reason, on the one hand,
neutrino oscillations can provide a direct window to probe physics beyond the
standard model~\cite{Smirnov:03}. On the other hand, some theoretical studies
of mixing in the context of quantum field
theory (QFT) by Blasone and Vitiello (BV)~\cite{BV:AP95,BlasoneP:03}
show the mixing problem may be related to more fundamental issues such as
unitarily inequivalent representations and the vacuum structure, and its study
is theoretically interesting for its own sake.

Nevertheless, the simpler plane-wave quantum mechanical
descriptions~\cite{Pontecorvo,PDG} seemed to be in accordance, in certain
realistic limits, with more refined descriptions, including various ingredients,
such as localization aspects~\cite{Kayser:81,Giunti:91,ccnishi04}, flavor
current densities~\cite{Zralek:98}, influence of creation and
detection processes~\cite{KiersWeiss,Giunti:wp:coh}, time-dependent
perturbation theory~\cite{Rich}, and intermediate neutrinos with path
integrals~\cite{Field}.
Moreover, many treatmens within the quantum field theory (QFT) framework were
also proposed~\cite{Giunti:93,Campagne,Beuthe:PRep,Kobzarev,GS,Dolgov,
Cardall.00}, aiming to solve the various unclear aspects of
the quantum mechanics of neutrino oscillations~\cite{Rich,Zralek:98}.

It has been known for a long time that the coherence necessary for neutrino
oscillations depends crucially on localization aspects of the particles involved
in the production of neutrinos~\cite{Kayser:81}. This point of view can be
supported by QFT arguments~\cite{GS} as well. It raises then the question of
how the  coherent superposition of mass eigenstate neutrinos, which is called a
``flavor'' eigenstate, is created~\cite{Dolgov}.
One way that became customary to avoid the ambiguities involving the question
on how neutrinos are created and detected is to use an {\it external} (E) wave
packet (WP) approach~\cite{Beuthe:PRep}, in contrast to an {\it intermediate}
(I) WP approach.
According to Ref.~\cite{Beuthe:PRep}, the IWP treatments are the simpler first
quantized ones treating the propagation of neutrinos as free localized wave
packets. In contrast, EWP approaches consider localized wave packets for the
sources and detection particles while the neutrinos were considered intermediate
virtual particles.
The central issue distinguishing the general IWP and EWP approaches is: despite
its direct unobservability, is the intermediate neutrino a real (on-shell)
particle propagating freely?
If the answer is affirmative, the IWP approaches would be a
good approximation of the oscillation phenomena.

On the other hand, another classification scheme can be used do classify the
various existing treatments considering a more physical criterion irrespective
of the use of WPs. Consider the descriptions of neutrino oscillations that
(A) include explicitly the interactions responsible for the mixing and those
(B) that only treat the propagation of neutrinos, i.e,. the mixing is an {\it
ad hoc} ingredient. A more subtle aspect in between would be the (explicit or
phenomenologically modelled) consideration of the production (and detection)
process(es).
In general, the IWP approaches are of type (B). The EWP approaches are of type
(A). The BV approach, although in the QFT formalism, is of type (B) since mixing
is introduced without explicitly including the interaction responsible for it.
The type (B) approaches have the virtue that they can be formulated in a way in
which total oscillation probability in time is always conserved and normalized
to one~\cite{ccnishi04,BV:AP95}. This feature will be present in all first
quantized approaches treated here (secs.\;\ref{sec:D} and \ref{sec:ST}) and in a
second quantized version (sec.\;\ref{sec:qft2}). If different observables are
considered, or a modeling of the details of the production and detection
processes is attempted, further normalization is
necessary~\cite{Giunti:91,BlasoneP:03,Giunti:wp:coh}. In such cases, the
oscillating observable might differ from the oscillation probability. On the
other hand, type (A) approaches tend to be more realistic and can account for
the production and detection processes giving experimentally observable
oscillation probabilities~\cite{Cardall.00}. Of course, they are essential to
the investigation of how neutrinos are produced and
detected~\cite{KiersWeiss,Dolgov}. We are not directly interested in these
matters here.

Considering first quantized type (B) approaches, some recent works treating the
flavor oscillation for spin one-half
particles~\cite{Bernardini:Euro,Bernardini:PRD05} have already find additional
oscillatorial effects compared to usual oscillation formulas with
WPs~\cite{Giunti:91,ccnishi04}. These effects are investigated and it is shown
in sec.\;\ref{sec:D} how these additional oscillatorial behavior, which have
characteristic frequencies much greater than usual oscillation frequencies,
comes from the interference between positive and negative frequency components
of the initial WP. It can be understood as a consequence of the impossibility to
simultaneously exclude all negative energy contributions of the initial
spinorial wave function with respect to bases characterized by different masses.
Moreover, this rapid oscillations are always present, independently of the
initial WP, if a well defined flavor is attributed to the initial WP.

To make clear the origin of the additional oscillatory contributions,
we calculate, in sec.~\ref{sec:ST}, the oscillation formula for a
charged spin 0 particle in the Sakata-Taketani Hamiltonian formalism~\cite{FV},
which is equivalent to the Klein-Gordon scalar wave equation.
(The explicit analysis with mixed Klein-Gordon equation is made in
Ref.~\onlinecite{Dvornikov}, paying special attention to the relativistic
initial value problem.) The oscillation formula in this case also possesses the
additional interference terms between positive and negative frequency parts,
very similar to the one obtained in the spin 1/2 case.
From this example we will see that these interference terms are inevitable from
a relativistic classical field theory perspective where covariance and causality
is required. It is not specially associated to the spin degree of freedom.

Another particular ingredient of neutrino oscillations can be included
naturally within Dirac theory:  the left-handedness of neutrinos created and
detected through weak interactions. This fact, for a Dirac
neutrino~\cite{Mohapatra}, implies an additional probability loss due to
conversion of left-handed neutrinos into right-handed neutrinos, which is
possible because chirality is no longer a constant of motion for massive Dirac
particles~\cite{chiral}. Although previous calculations~\cite{Bernardini:PRD05}
have shown an approximate contribution to this effect, we calculate in
sec.~\ref{subsec:D:L} the complete effect.

Concerning type (A) approaches, specifically the EWP description, we are
interested to analyze further how is the propagation of intermediate virtual
neutrinos. The framework where the investigations on first quantized approaches
are made here is based on the calculation of the evolution kernels for free
theories in presence of mixing. This enable us to deduce general oscillation
probabilities in which there is explicit decoupling from the oscillating  part
(where all the oscillation information rests) and the initial wave packet.
Another advantage of doing the calculations this way is that it resembles
the propagator methods in covariant perturbation theory, which EWP approaches
are based on. The free evolution kernel for fermions have a close relationship
with the Feynman propagator used in QFT. What is common to both is that both
particle and antiparticle parts contribute to the evolution or propagation. The
necessity of the negative frequency part in the free evolution kernel is
required from completeness and causality arguments but it also leads to the
interference of positive and negative frequencies in flavor oscillation, treated
in secs.~\ref{sec:D} and \ref{sec:ST}. Then the question also arises in EWP
approaches: are there contributions from both particles and antiparticles in the
propagation of virtual neutrinos? In a simple microscopic scattering process,
this question is meaningless since virtual particles are usually off-shell
particles and must naturally have both contributions. However in EWP approaches
the neutrinos propagate through macroscopic distances and, indeed, it can be
shown~\cite{GS,Dolgov} that the virtual neutrinos are on-shell particles.
There is no discussion, though, about the possibility of neutrino and
antineutrino contributions to the process;  both can be on-shell.
This investigation is carried on in sec.~\ref{sec:qft} calculating explicitly
the amplitude of production/propagation/detection process in an EWP approach.

As a last task, we develop a simple, type (B), second quantized description
of flavor oscillation in sec.~\ref{sec:qft2} using the free second quantized
spin 1/2 fermionic theory in presence of mixing.
This treatment has some similarities with the BV formalism but it does not
require the introduction of flavor Fock spaces and Bogoliubov transformations.
It means that the Fock space considered will be the one spanned by the mass
eigenstates. Within this formalism it will be shown that the additional rapid
oscillation contributions calculated through first quantized approaches do not
survive the second quantization since only superpositions of particles
(antiparticles) are used as initial neutrino (antineutrino) ``flavor'' states.
Moreover, this property is not satisfied in the BV approach because the BV
flavor states are mixtures of particle and antiparticle components; this is the
ingredient responsible for a different oscillation probability~\cite{BV:AP95}.

\section{Flavor Oscillation for Dirac fermions}
\label{sec:D}

It is well known that the Dirac equation can give a significantly good
description of a Dirac fermion if its inherent localization is much bigger
than its Compton wave length; usually this is associated with weak external
fields. For example, the spectrum for the hydrogen atom can be obtained with the
relativistic corrections included (fine structure)\,\cite[\footnotesize
p.~72]{IZ}. One of the terms responsible for fine structure, the Darwin term,
can be interpreted as coming from the interference between positive and negative
frequency parts ({\it zitterbewegung}) of the hydrogen eigenfunction in Dirac
theory compared to the nonrelativistic theory\;\cite{FV}. On the other hand a
situation where the theory fails to give a satisfactory physical description is
exemplified by the Klein paradox\,\cite[\footnotesize p. 62]{IZ}: the
transmission coefficient for a electron moving towards a step barrier becomes
negative for certain barrier heights, exactly when the localization of the
electron wave function inside the barrier is comparable with its Compton wave
length.

Bearing in mind that first quantized approaches may fail under certain
conditions we will treat in this section the flavor oscillation problem using
the free Dirac theory in presence of two families mixing.
The extension to treat three families of neutrinos is straightforward.
A matricial notation will be used throughout the article for the first
quantized approaches to express the mixing.

In matricial notation the mixing relation between flavor wave functions
${\Psi}_f^{\TT}(\bx)\equiv (\psi_{\nu_e}^{\TT}(\bx),\psi_{\nu_\mu}^{\TT}(\bx))$
and mass wave functions ${\Psi}_m^{\TT}(\bx)\equiv (\psi_1^{\TT}(\bx),
\psi_2^{\TT}(\bx))$  is
\eq{
\label{mixing}
\Psi_{f}(\bx)
\equiv
U\Psi_{m}(\bx)
=
\left(\begin{array}{cc}
\phantom{\!\!\!-}\cos\!\theta & \sin\!\theta \cr
\!\!\!-\sin\!\theta & \cos\!\theta
\end{array}\right)
\Psi_{m}(\bx)
~.
}
Each mass wave function is defined as a four-component
spinorial function $\psi_n(\bx,t)$, $n=1,2$ that satisfy the free Dirac
equation
\eq{
\label{nu:n:eq}
i\frac{\partial}{\partial t}\psi_n(\bx,t)=H_n^D\psi_n(\bx,t)~,~~n=1,2~,
}
where the free Hamiltonian is the usual
\eq{
\label{nu:n:H}
H_n^D\equiv -i\bs{\alpha}.\nabla + \beta m_n~,~~n=1,2~.
}

We will work in the flavor diagonal basis. This choice defines the flavor
basis vectors simply as
\eq{
\label{versor}
\hat{\nu}_e^{\TT}=(1,0)~, ~~
\hat{\nu}_{\mu}^{\TT}=(0,1)~,
}
while the flavor projectors are obviously
\eq{
\label{proj:f}
\mathsf{P}_{\nu_{\alpha}}\equiv \hat{\nu}_{\alpha}\hat{\nu}_{\alpha}^{\TT}
\,.
}
Actually, as an abuse of notation, the equivalence $U \sim U\otimes \id_{D}$ is
implicit , as well as, $\mathsf{P}_{\nu_\alpha}\sim
\mathsf{P}_{\nu_\alpha}\otimes \id_D$; the symbol $\id_D$ refers to the identity
matrix in spinorial space.

The total Hamiltonian governing the dynamics of $\Psi_m$ is
$H^D=\mathrm{diag}(H_1^D,H_2^D)$. From the considerations above, $\Psi_f(\bx,t)$
satisfy the equation \eq{
\label{nu:f:eq}
i\frac{\partial}{\partial t}\Psi_f(\bx,t)\equiv
U H^D U^{-1} \Psi_f(\bx,t)~.
}
The solution to the equation above can be written in terms of a flavor evolution
operator $K^D$ as
\eq{
\label{f:evol:D}
\Psi_f(\bx,t)=K^D(t)\Psi_f(\bx,0)=
\int\!\! d^3\bx^{\pr}\,K^D(\bx-\bx^{\pr};t)\Psi_f(\bx^{\pr},0)
~,
}
where
\eq{
\label{KD}
K^D(\bx-\bx^{\pr};t)
=\int\!\frac{d^3\bp}{(2\pi)^3}\, K^D(\bp;t)\,e^{i\bp.(\bx-\bx^{\pr})} ~.
}
We can calculate $K^D(t)$ in any representation (momentum or position) as
\eqarr{
\label{KD:mtrx}
K^D(t)
&=&
U e^{-iH^Dt}U^{-1}
\cr
&=&
\left(
\begin{array}{cc}
\cos^2\!\theta\,e^{-iH_1^Dt}+\sin^2\!\theta\,e^{-iH_2^Dt} &
-\cos\!\theta\sin\!\theta(e^{-iH_1^Dt}-e^{-iH_2^Dt}) \cr
-\cos\!\theta\sin\!\theta(e^{-iH_1^Dt}-e^{-iH_2^Dt}) &
\sin^2\!\theta\,e^{-iH_1^Dt}+\cos^2\!\theta\,e^{-iH_2^Dt}
\end{array}
\right)~.
}
The conversion probability is then
\eqarr{
\label{prob:em:D}
\mathscr{P}(\ms{\nu_e\!\rightarrow\!\nu_\mu};t)
&=&
\int\! d\bx\,
\Psi_f^{\dag}(\bx,0)K^{D\,\dag}(t)
\mathsf{P}_{\nu_\mu}
K^D(t)\Psi_f(\bx,0)
\cr
&=&
\int\! d\bp\,
\tilde{\psi}_{\nu_e}^{\dag}(\bp)(K_{\mu e}^{D})^\dag
K_{\mu e}^D(\bp,t) \tilde{\psi}_{\nu_e}(\bp)
~,
}
satisfying the initial condition $\Psi_f^{\TT}(\bx,0)=(\psi^{\TT}_{\nu_e}(\bx,0)
, 0)$. Such initial condition implies, in terms of mass eigenfunctions,
$\psi_1(\bx,0)=\psi_2(\bx,0)=\psi_{\nu_e}(\bx)$, as a requirement to obtain an
initial wave function with definite flavor~\cite{ccnishi04}. The function
$\tilde{\psi}_{\nu_e}(\bp)$ denotes the inverse Fourier transform of
$\psi_{\nu_e}(\bx)$ (see Eqs.~\eqref{fourier:xp} and \eqref{fourier:px}).

Before obtaining the conversion probability for Dirac fermions, let us replace
the spinorial functions $\psi_n(\bx)$ by spinless one-component wave functions
$\varphi_n(\bx)$ in the flavor wave function ${\Psi}_f^{\TT}(\bx)\rightarrow
(\varphi_{\nu_e}(\bx),\varphi_{\nu_\mu}(\bx))$ and mass wave function
${\Psi}_m^{\TT}(\bx)\rightarrow (\varphi_1(\bx), \varphi_2(\bx))$. We also
replace  the Dirac Hamiltonian in momentum space $H^D_n(\bp)$ \eqref{nu:n:H} by
the relativistic energy $E_n(\bp)=\sqrt{\bp^2+m^2_n}$.
Inserting these replacements into Eq.\,\eqref{prob:em:D} we can recover the
usual oscillation probability~\cite{ccnishi04,Bernardini:PRD05}
\eqarr{
\label{prob:em:S}
\mathscr{P}(\ms{\nu_e\!\rightarrow\!\nu_\mu};t)&=&
\int\! d\bx |\hat{\nu}_\mu^{\TT}\Psi_{f}(\bx,t)|^2
\cr
&=&
\int\! d\bp\,|K^S_{\mu e}(\bp,t)\tilde{\varphi}_{\nu_e}(\bp)|^2
\cr
&=&
\int\!d\bp\mathscr{P}(\bp,t)
|\tilde{\varphi}_{\nu_e}(\bp)|^2
~,
}
where $\Psi_{f}(\bx,0)^{\TT}=(\varphi_{\nu_e}(\bx)^\TT,0)$,
$K^S_{\mu e}(\bp,t)\equiv(K^S)_{21}=
-\sin\theta\cos\theta(e^{-iE_1(\bp)t}-e^{-iE_2(\bp)t})$ and
\eq{
\label{P}
\mathscr{P}(\bp,t)=\sin^2\!2\theta\sin^2\!(\Delta E\mfn{(\bp)}t/2)
~
}
is just the standard oscillation formula. The conversion probability
\eqref{prob:em:S} in this case is then the standard oscillation probability
smeared out by the initial momentum distribution. If the substitution
$|\tilde{\varphi}_{\nu_e}(\bp)|^2\rightarrow \delta^3(\bp-\bp_0)$ is made
the standard oscillation formula is recovered: it corresponds to the plane-wave
limit.

After we have checked the standard oscillation formula can be recovered for
spinless particles restricted to positive energies in the plane-wave limit, we
can return to the case of Dirac fermions.
We can obtain explicitly the terms of the mixed evolution
kernel \eqref{KD:mtrx} by using the property of the Dirac Hamiltonian in
momentum space ${H_n^D}^2=E_n^2(\bp)\,\id_D$, which leads
\eqarr{
\label{KD:em:0}
(K_{\mu e}^{D})^\dag K_{\mu e}^D(\bp,t)
&=&
\label{KD:em}
\mathscr{P}(\bp,t)[1-f(\bp)]\id_D
\cr&&
+\, \sin^2\!2\theta f(\bp)\sin^2(\bar{E}t)\id_D
~, }
where
\eq{
\label{f}
f(\bp)=\frac{1}{2}[1-\frac{\bp^2+m_1m_2}{E_1E_2}]~,
}
and $\mathscr{P}(\bp,t)$ is the standard conversion probability function
\eqref{P}. A unique implication of Eq.~\eqref{KD:em}, which is
proportional to the identity matrix in spinorial space, is that the conversion
probability \eqref{prob:em:D} does not depend on the spinorial structure of
the initial flavor wave function but only on its momentum density as
\begin{align}
\label{prob:em:D2}
\mathscr{P}(\ms{\nu_e\!\rightarrow\!\nu_\mu};t)=
\int\! d\bp\,\{\mathscr{P}(\bp,t)[1-f(\bp)] \hs{3em}
\nonumber\\
+~\sin^2\!2\theta f(\bp)\sin^2(\bar{E}t)\}\,
\tilde{\psi}^\dag_{\nu_e}(\bp)\tilde{\psi}_{\nu_e}(\bp)
\,.
\end{align}
(The tilde will denote the inverse Fourier transformed function throughout
this paper.) Furthermore,  the modifications in Eq.~\eqref{prob:em:D2} compared
to the scalar conversion probability \eqref{prob:em:S} are exactly the same
modifications found in Ref.~\cite{Bernardini:Euro} and
Ref.~\cite{Bernardini:PRD05} after smearing out through a specific
gaussian wave packet.

The conservation of total probability
\eq{
\label{prob:cons}
\mathscr{P}(\ms{\nu_e\!\rightarrow\!\nu_\mu};t)+\mathscr{P}(\ms{\nu_e\!
\rightarrow\!\nu_e};t)=1
~,
}
is automatic in virtue of
\eq{
\label{prob:cons:D}
K_{ee}^{D\dag}(\bp,t)K_{ee}^D(\bp,t)+
K_{\mu e}^{D\dag}(\bp,t)K_{\mu e}^D(\bp,t)=\id_D
~,
}
and the survival and conversion probability for an initial muon neutrino are
identical to the probabilities for an initial electron neutrino because of the
relations
\eqarr{
\label{prob:me:D}
K_{\mu\mu}^{D\dag}(\bp,t)K_{\mu\mu}^D(\bp,t)&=&
K_{ee}^{D\dag}(\bp,t)K_{ee}^D(\bp,t)
~,
\\
K_{\mu e}^{D\dag}(\bp,t)K_{\mu e}^D(\bp,t)&=&
K_{e\mu}^{D\dag}(\bp,t)K_{e\mu}^D(\bp,t)
~.
}

To explain the origin of the additional oscillatory terms in
Eq.~\,\eqref{prob:em:D2} it is instructive to rewrite the free Dirac time
evolution operator, in momentum space, in the form
\eq{
\label{UD}
e^{-iH_n^Dt}=e^{-iE_nt}\Lambda_{n+}^D+e^{iE_nt}\Lambda_{n-}^D
~,
}
where
\eq{
\label{Lambda+-}
\Lambda_{n\pm}^D=\frac{1}{2}(\id_D\pm \frac{H_n^D}{E_n})
~,
}
are the projector operators to positive (+) or negative (-) energy eigenstates
of $H_n^D$. By using the decomposition above \eqref{UD}, we can analyze
$K^D_{\mu e}$ in Eq.~\eqref{KD:mtrx}, which contains the terms
\eqarr{
e^{iH_1^Dt}e^{-iH_2^Dt}&=&e^{i\Delta Et}\Lambda_{1+}^D\Lambda_{2+}^D
+e^{-i\Delta Et}\Lambda_{1-}^D\Lambda_{2-}^D
\cr &&
+~e^{i2\bar{E}t}\Lambda_{1+}^D\Lambda_{2-}^D
+e^{-i2\bar{E}t}\Lambda_{1-}^D\Lambda_{2+}^D
\,,~~~
}
plus its hermitian conjugate.
Since $\Lambda_{1\pm}^D\Lambda_{2\mp}^D\neq 0$, it can be seen that the rapid
oscillating terms come from the interference between, e.g., the positive
frequencies of the Hamiltonian $H_1^D$ and negative energies of the Hamiltonian
$H_2^D$. One may think that by restricting the initial wave function to contain
only positive energy contributions would eliminate the rapid oscillatory terms,
as {\it zitterbewegung} disappears for superpositions of solely positive  energy
states in Dirac theory~\cite{IZ}, but it does not happen.
The positive energy eigenfunctions with respect to a basis characterized by a
mass $m_1$ necessarily have non-null components of negative energy with respect
to another basis characterized by $m_2$ (this point is illustrated in appendix
\ref{app:dec}). Thus the rapid oscillatory contributions are an inevitable
consequence of this framework and it is always present independently of the
initial WP, if initially a definite flavor is chosen.
However, its influence, quantified by the function $f(\bp)$ in
Eq.~\,\eqref{f}, is negligible for momentum distributions around
ultra-relativistic values~\cite{Bernardini:Euro}. This rapid oscillatory terms
will also be found for charged spin 0 particle oscillations in the next section,
with contributions slightly different from the ones obtained for spin 1/2
particles.

\subsection{Inclusion of Left-Handedness}
\label{subsec:D:L}

Until this point, we have been considering the oscillation of general flavor
``particle number'' for general Dirac neutrinos.
However, due to the left handed nature of weak interactions
 only left-handed components are produced and detected.
To incorporate this fact into, for example, the conversion probability in
Eq.~\eqref{prob:em:D}, it is sufficient to use initial left-handed WPs and
replace the kernel of Eq.~\eqref{KD:em:0} by the projected counterpart
\begin{align}
\label{KD:em:L}
LK_{\mu e}^{D\dag}(\bp,t)LK_{\mu e}^D(\bp,t)L
=
\mathscr{P}^D(\bp,t)L
\hs{6em}
\nonumber \\
-\frac{1}{4}\sin^22\theta\bigg(
\frac{m_1}{E_1}\sin E_1t-\frac{m_2}{E_2}\sin E_2t
\bigg)^2L
~,~~
\end{align}
where $\mathscr{P}^D(\bp,t)=K_{\mu e}^{D\dag}(\bp,t)K_{\mu e}^D(\bp,t)$ is the
conversion kernel of Eq.~\eqref{KD:em} and $L=(1-\gamma_5)/2$ is the projector to
left chirality. The conservation of total probability \eqref{prob:cons} no
longer holds because there is a probability loss due to the undetected right
handed component
\eq{
\label{KD:em:R}
LK_{\mu e}^{D\dag}RK_{\mu e}^D(\bp,t)L
=
\frac{1}{4}\sin^22\theta\bigg(
\frac{m_1}{E_1}\sin E_1t-\frac{m_2}{E_2}\sin E_2t\!\bigg)^2\!L
~,
}
where $R=(1+\gamma_5)/2$ is the projector to right chirality. We can see that
the probability loss \eqref{KD:em:R} is proportional to the ratio $m_n^2/E_n^2$
which is negligible for ultra-relativistic neutrinos.
The total probability loss for an initial left-handed electron neutrino
turning into right-handed neutrinos, irrespective of the final flavor, is given
by the kernel
\eq{
\label{KD:LR}
LK_{\mu e}^{D\dag}RK_{\mu e}^D(\bp,t)L
+
LK_{ee}^{D\dag}RK_{ee}^D(\bp,t)L
=
\Big[\cos^2\!\theta \Big(\frac{m_1}{E_1}\Big)^2\sin^2E_1t
+\sin^2\!\theta \Big(\frac{m_2}{E_2}\Big)^2\sin^2E_2t\Big]L
~.
}

To obtain the unphysical complementary kernels responsible for the conversion
of right-handed component to right-handed and left-handed components, it is
enough to make the substitution $L\leftrightarrow R$ in all formulas.

\section{Flavor Oscillation for Spin 0}
\label{sec:ST}

The derivation of the usual conversion probability \eqref{prob:em:S} takes into
account only the positive frequency contributions.
The mass wave function used to obtain Eq.\,\eqref{prob:em:S} corresponds to the
solutions of the wave equation
\eq{
\label{we:S}
i\frac{\partial}{\partial t}\varphi(\bx,t)
=\sqrt{-\nabla^2+m^2}\,\varphi(\bx,t)~,
}
which is equivalent to the Dirac equation in the Foldy-Wouthuysen
representation~\cite{FW}, restricted to positive energies.
The evolution kernel for this equation is not satisfactory from the point of
view of causality~\cite[p.18]{Thaller}, i.e, the kernel is not null for spacelike
intervals. Moreover, the eigenfunctions restricted to one sign of energy do not
form a complete set~\cite{FV}.

To recover a causal propagation in the spin 0 case, the Klein-Gordon wave
equation must be considered. In the first quantized version, the spectrum of
the solutions have positive and negative energy as in the Dirac case.
However, to take advantage of the Hamiltonian formalism used so far, it is more
convenient do work in the Sakata-Taketani (ST) Hamiltonian formalism~\cite{FV}
where each mass wave function is formed by two components
\eq{
\label{Phi}
\Phi_n(\bx,t)=
\left(
\begin{array}{c}
\varphi_n(\bx,t) \cr \chi_n(\bx,t)
\end{array}
\right)
~,~~n=1,2\,.
}
The components $\varphi$ and $\chi$ are combinations of the usual scalar
Klein-Gordon wave function $\phi(x)$ and its time derivative
$\partial_0\phi(x)$. This is necessary since the Klein-Gordon equation is
a second order differential equation in time and the knowledge of the function
and its time derivative is necessary to completely define the time evolution.

The time evolution in this formalism is governed by the Hamiltonian~\cite{FV}
\eq{
\label{H:ST}
H_n^{\mt{ST}}=-(\tau_3+i\tau_2)\frac{\nabla^2}{2m_n}+m_n^2
~,
}
which satisfies the condition $(H_n^{\ST})^2=(-\nabla^2+m_n^2)\id_{\ST}$,
like the Dirac Hamiltonian \eqref{nu:n:H}. The $\tau_k$ represents the usual
Pauli matrices and $\id_{\ST}$ is the identity matrix.

A charge density~\cite{endnote1} can be defined as
\eq{
\label{rho:S}
\bar{\Phi}_n\Phi_n \equiv \Phi_n^\dag\tau_3\Phi_n=|\varphi_n|^2-|\chi_n|^2~,
}
which is equivalent to the one found in Klein-Gordon notation
$i\phi^*\overset{\leftrightarrow}{\partial}_0\phi$. Needless to say, this
density \eqref{rho:S} is only non-null for complex (charged) wave functions.
The charge density $\bar{\Phi}\Phi$ is the equivalent of fermion probability
density $\psi^{\dag}\psi$ in the Dirac case, although the former is not positive
definite as the latter. The adjoint $\bar{\Phi}=\Phi^\dag\tau_3$ were defined to
make explicit the (non positive definite) norm structure of the conserved charge
\eq{
\label{Q:ST}
\int\! d\bx\,\bar{\Phi}_n(\bx,t)\Phi_n(\bx,t) \equiv
(\Phi_n,\Phi_n)=\text{time independent}
~.
}
Consequently, the adjoint of any operator $\Omega$ can be defined as
$\bar{\Omega}=\tau_3\Omega^\dag\tau_3$, satisfying
$(\bar{\Omega}\Phi,\Phi)=(\Phi,\Omega\Phi)$. Within this notation, the
Hamiltonians of Eq.~\eqref{H:ST} is self-adjoint, $\bar{H}_n^{\ST}=H_n^{\ST}$,
and the time invariance of Eq.~\eqref{Q:ST} is assured.

We can assemble, as in the previous section, the mass wave functions into
$\Psi_m^\TT \equiv (\Phi_1^\TT,\Phi_2^\TT)$ and the flavor wave functions into
$\Psi_f^\TT \equiv (\Phi_{\nu_e}^\TT,\Phi_{\nu_\mu}^\TT)$, satisfying
the mixing relation $\Psi_f\equiv U\Psi_m$. The equivalence of
$U\sim U\otimes \id_{\ST}$ and of $\mathsf{P}_{\nu_\alpha}
\sim \mathsf{P}_{\nu_\alpha}\otimes \id_{\ST}$ are implicit without
modification in the notations. Then, the time evolution of $\Psi_f$ can be given
through a time evolution operator $K^{\ST}$ acting in the same form as in
Eq.~\eqref{f:evol:D}.
In complete analogy to the calculations from Eq.~\eqref{KD}
to Eq.~\eqref{prob:em:D}, we can define the conversion probability as
\eqarr{
\label{prob:em:ST}
\mathscr{P}(\ms{\nu_e\!\rightarrow\!\nu_\mu};t)
&=&\int\! d\bx\,
\bar{\Psi}_f(\bx,0)\overline{K^{\ST}(t)}\mathsf{P}_{\nu_\mu}K^{\ST}(t)
\Psi_f(\bx,0) \cr &=&
\int\! d\bp\,\overline{\tilde{\Phi}_e(\bp)}\overline{K_{\mu e}^{\ST}}
K_{\mu e}^{\ST}(\bp,t)\tilde{\Phi}_e(\bp)
~,
}
where $\Psi_f(\bx,0)^\TT=(\Phi_e(\bx)^\TT,0)$. The adjoint operation were
also extended to $\bar{\Psi}_f=\Psi_f^\dag(\id_{\theta}\otimes\tau_3)$, where
$\id_\theta$ is the identity in mixing space.

The information of time evolution, hence oscillation, is all encoded in
\eqarr{
\label{KST:em:0}
\overline{K_{\mu e}^{\ST}}K_{\mu e}^{\ST}(\bp,t)
&=&
\label{KST:em}
\mathscr{P}(\bp,t)[1-f(\mu\bp)]\id_{\ST}
\cr &&
+~\sin^2\!2\theta f(\mu\bp)\sin^2(\bar{E}t)\id_{\ST} ~,
}
where the function $f(\bp)$ were already defined in Eq.~\eqref{f} and
\eq{
\label{mu}
\mu=\sqrt{\frac{1}{2}(\frac{m_1}{m_2}+\frac{m_2}{m_1})}
~.
}
The factor $\mu\ge 1$ determines the difference with the Dirac case in
Eq.~\eqref{KD:em}.
The equality $\mu=1$ holds when $m_1=m_2$, i.e., when there is no oscillation.

\section{Connection with quantum field theory}
\label{sec:qft}

The main improvement of the covariant approaches developed in secs.\;\ref{sec:D}
and \ref{sec:ST} is that the propagation kernels governed by Dirac and
Sakata-Taketani Hamiltonians are causal, i.e., are null for spacelike
separations (see Eqs.~\eqref{S:D} and \eqref{S:ST} and
Refs.~\cite{Thaller,IZ,Roman}). On the contrary, the kernel of spinless
particles restricted only to positive energies is not null for spacelike
intervals~\cite{Thaller}. From the point of view of relativistic classical field
theories, a causal kernel guarantees, by the Cauchy theorem, the causal
connection between the wave-function in two spacelike surfaces at different
times~\cite{Roman}.

To compare the IWP and EWP approaches it
is useful to rewrite the Dirac evolution kernel for a fermion of mass $m_n$,
present in Eq.~\eqref{f:evol:D}, in the form~\cite[\small p.89]{IZ}
\eqarr{ \label{S}
K_n^D(x-y)&=&
\underset{s}{\textstyle \sum}\int\!
\frac{d^3\!p\,}{2E_n}
[u^s_n(x;\!\bp)\bar{u}^s_n(y;\bp)+v^s_n(x;\!\bp)\bar{v}^s_n(y;\bp)]\gamma_0
\cr
&\equiv&iS(x-y;m_n)\gamma_0
~,~~n=1,2,
}
where $(x-y)^0=t, (x-y)^i=(\bx-\bx')^i$ when compared to the notation
of Eq.~\eqref{f:evol:D}.
The spinorial functions $u,v$, are the free solutions of the Dirac equation and
they are explicitly defined in appendix \ref{app:def}. (More familiar forms
for the function $S$ are also shown in appendix \ref{app:def}.)
Clearly the function $iS(x-y;m_n)=\bra{0}\{\psi_n(x),\bar{\psi}_n(y)\}\ket{0}$
satisfies the homogeneous Dirac equation with mass $m_n$ \eqref{nu:n:eq} and it
is known to be null for spacelike intervals $(x-y)^2<0$~\cite{Thaller,Roman}.

In contrast, the Feynman propagator $iS_F(x-y)$ appears in QFT. It is a
Green function for the inhomogeneous Dirac equation obeying particular
boundary conditions. The EWP approaches use this Green function for the
propagation of virtual neutrinos. To directly compare the Feynman propagator to
the kernel in Eq.~\eqref{S} we can write $iS_F$ in the form
\eqarr{
\label{S:F}
iS_F(x-y;m_n)&\equiv &
\bra{0}T(\psi_n(x),\bar{\psi}_n(y))\ket{0}
\cr
&=&
\underset{s}{\textstyle \sum}\int\!
\frac{d^3\!p}{2E_n}\,
[u^s_n(x;\!\bp)\bar{u}^s_n(y;\bp)\theta(x_0-y_0)
\cr
&&\phantom{\sum_s\int\!\frac{d^3\!p}{2E_n}\,}\hs{-1.9ex}
- v^s_n(x;\!\bp)\bar{v}^s_n(y;\bp)\theta(y_0-x_0)]
~.
}
Although the function $S_F$ is called causal propagator, it is not null for
spacelike intervals, and it naturally arises in QFT when interactions are present
and treated in a covariant fashion.
Equation~\eqref{S:F} shows that the propagator $S_F$
describes positive energy states propagating forward in time and negative energy
states propagating backward in time~\cite[\small p.91]{IZ}. At a first glance,
both neutrino and antineutrino parts of Eq.~\eqref{S:F} seem to contribute to
the space-time integrations present in covariant perturbation theory, as
neutrino-antineutrino contributions in Eq.~\eqref{S} have led to
Eq.~\eqref{prob:em:D2}.

In the following we will show in an EWP approach that for large separations
between production and detection both neutrino and antineutrino
parts may contribute as intermediate neutrinos for certain situations.

We will follow the calculations made in Ref.\,\cite{Dolgov}, using, instead of
the scalar interaction, the effective charged-current weak lagrangian
\eqarr{
\label{LW}
\mathscr{L}_W&=&
G\sum_{i,\alpha=1}^{N=3}
\,[
\bar{l}_\alpha (x)\gamma^\mu L\,U_{\alpha i}\nu_i(x)J_\mu(x)
\cr&&\hs{3em}
+
\bar{\nu}_i(x)U^*_{\alpha i}\gamma^\mu L\, l_\alpha (x)J_\mu^\dag(x)
\,]
\\
&=&\mathscr{L}_1+\mathscr{L}_1^\dag
~,
}
where $G=\sqrt{2}G_F$ and
$J_\mu$ is the sum of any effective leptonic or hadronic current. The
lagrangian \eqref{LW} is written only in terms of physical mass eigenstate
fields, which coincides with flavor eigenstate fields only for the charged
leptons: $l_1(x)\equiv e(x), l_2(x)\equiv \mu(x), \ldots$\,.

Suppose the process\,\cite{Dolgov,Giunti:qft:02} where a charged lepton $l_\alpha $
hit a nucleus A turning it into another nucleus A$'$ with emission of a
neutrino (this process happens around $x_A$). Subsequently the neutrino
travels a long distance and hit a nucleus B which transforms into B$'$ emitting
a lepton $l_\beta $ (this process happens around $x_B$). The whole process looks like
$l_\alpha +A+B\rightarrow l_\beta +A'+B'$ with transition amplitude given by
\eq{
\label{A:1}
\bra{A'(\bp'_A),B'(\bp_B'),l_\beta (\bp_\beta )}S
\ket{A,B,l_\alpha }
~.
}
The final states are momentum eigenstates while the initial states are
localized\,\cite{Dolgov}. The lowest order nonzero contribution of the
scattering matrix $S$ to Eq.\;\eqref{A:1} is second order in the lagrangian
\eqref{LW}. More explicitly, the term that contributes to the amplitude
\eqref{A:1} comes from \eqarr{
S^{(2)}&=&\frac{i^2}{2}T\aver{\lag_W}^2
=-\frac{1}{2}T\aver{\lag_1+\lag_1^\dag}^2
\\
\label{S2:1}
&\sim&-T\aver{\lag_1}\aver{\lag_1^\dag}
\\
\label{S2:2}
&\sim&
-G^2\int d^4x d^4y \sum_{\beta\alpha}\lag_{\beta\alpha}(x,y)
~,
}
where $\aver{~}$ stands for space-time integration and
\eq{
\lag_{\beta\alpha}(x,y)
\equiv
\sum_i
\!:\!J_\mu(x)\bar{l}_\beta (x)\gamma^\mu L U_{\beta i}\,iS_{F}(x-y;m_i)U^*_{\alpha
i} \gamma^\nu L l_\alpha (y)J_\nu^\dag(y)\!:
~.
}
In Eq.\,\eqref{S2:1} we kept only the mixed product and in Eq.\,\eqref{S2:2} we
kept from all possible terms in Wick expansion\,\cite[\mfn{\rm p.180}]{IZ} only
the term responsible for the transition of interest.

Then the transition amplitude \eqref{A:1} can be calculated as
\begin{align}
-G^{-2}\bra{A'(\bp'_A),&B'(\bp_B'),l_\beta (\bp_\beta )}S^{(2)}\ket{A,B,l_\alpha }
\cr&=
\int d^4y\,d^4x\,
\bra{B'(\bp_B')}J_\mu(y)\ket{B}
\bra{A'(\bp'_A)}J_\nu^\dag(x)\ket{A}
\cr&\hs{3em}\times
\bar{u}_\beta (y,\bp_\beta )\gamma^\mu L
\sum_iU_{\beta i}U^*_{\alpha i}iS_F(y-x;m_i)\gamma^\nu L\,
\bra{0}l_\alpha (x)\ket{l_\alpha }
\label{A:3}
\\
\label{Ai}
&\equiv \sum_i U_{\beta i}U^*_{\alpha i}\,\mcal{A}_i
~.
\end{align}
The initial states must be chosen in such a way that $A,l_\alpha $ are localized
around $x_A=(t_A,\bx_A)$ and $B$ is localized around $x_B=(t_B,\bx_B)$, since
we are ultimately interested in large separations $|\bx_B-\bx_A|$.
We can implement explicitly those localization conditions into the wave packets
\eqarr{
\label{J:B}
\bra{B'(\bp_B')}J_\mu(y)\ket{B}&=&\frac{1}{(2\pi)^{\frac{3}{2}}}
\int \widehat{d\bq_B}\,e^{ip'_B.y}J^{BB'}_\mu(\bq_B,\bp'_B)\psi_B(\bq_B)
e^{-iq_B.(y-x_B)}
\\
\label{J:A}
\bra{A'(\bp'_A)}J_\nu^\dag(x)\ket{A}&=&\frac{1}{(2\pi)^{\frac{3}{2}}}
\int \widehat{d\bq_A}\,e^{ip'_A.y}J^{AA'}_\mu(\bq_A,\bp'_A)\psi_A(\bq_A)
e^{-iq_A.(x-x_A)}
\\
\label{psi:j}
\bra{0}l_\alpha (x)\ket{l_\alpha }&=&\frac{1}{(2\pi)^{\frac{3}{2}}}
\int \widehat{d\bq_\alpha }
\,\psi_\alpha (\bq_\alpha )e^{-iq_\alpha .(x-x_A)}
~,
}
where $\widehat{d\bq}=d\bq(2E(\bq))^{-1/2}$,
$J^{BB'}_\mu(\bq_B,\bp'_B)=\bra{B'(\bp_B')}J_\mu(0)\ket{B(\bq_B)}$ and
$J^{AA'}_\nu(\bq_A,\bp'_A)=\bra{A'(\bp'_A)}J_\nu^\dag(0)\ket{A(\bq_A)}$.

Following the calculations from Eq.\,\eqref{A:3} with the localization
aspects of Eqs.\,\eqref{J:B}-\eqref{psi:j} included, we arrive at
\eqarr{
\mcal{A}_i&=&
\frac{1}{(2\pi)^{6}}
\int \widehat{d\bq_B}
\int \widehat{d\bq_A}
\int \widehat{d\bq_\alpha }
J^{BB'}_\mu(\bq_B,\bp'_B)\psi_B(\bq_B)
J^{AA'}_\mu(\bq_A,\bp'_A)\psi_A(\bq_A)
\cr&&
\label{A:4}
e^{iq_B.x_B}e^{i(q_A+q_\alpha ).x_A}
\bar{u}_\beta (\bp_\beta )\gamma^\mu L
\Big[
\int d^4x\,d^4y\,e^{i\kappa_\beta.y}e^{-i\kappa_\alpha .x}
iS_i(y-x)
\Big]
\gamma^\nu L\psi_\alpha (\bq_\alpha )
~,
}
where $\kappa_\beta=(\kappa^0_\beta ,\bs{\kappa}_\beta )$,
$\kappa_\alpha =(\kappa^0_\alpha ,\bs{\kappa}_\alpha )$ and
\begin{align}
\label{kk}
\bs{\kappa}_\beta  & \equiv \bp_\beta +\bp'_B-\bq_B~,
 \> \hs{2em}  \kappa^0_\beta  \> \equiv
E_\beta (\bp_\beta )+E_{B'}(\bp'_B)-E_B(\bq_B)~,
\\
\label{kj}
\bs{\kappa}_\alpha  & \equiv \bp_\alpha -\bp'_A+\bq_A~,
\> \hs{2em} \>\kappa^0_\alpha \> \equiv
E_\alpha (\bq_\alpha )-E_{A'}(\bp'_A)+E_A(\bq_A)~.
\end{align}

By using the results of Eqs.\,\eqref{wS:+} and \eqref{wS:-} the expression
between square brackets in Eq.\,\eqref{A:4} gives
\begin{align}
2\pi\delta(\kappa^0_\beta -\kappa^0_\alpha )
\int &d\bx\,d\by\,
\frac{-i}{4\pi r}e^{ik_\omega r}e^{-i\bs{\kappa}_\beta .\by}e^{i\bs{\kappa}_\alpha .\bx}
\cr&\times
[u_i(k_\omega\hat{\br})\bar{u}_i(k_\omega\hat{\br})\theta(\omega_i-m_i)
-v_i(-k_\omega\hat{\br})\bar{v}_i(-k_\omega\hat{\br})\theta(-\omega_i-m_i)
]
\label{[]}
~,
\end{align}
where $r\equiv |\by-\bx|$, $\hat{\br}\equiv (\by-\bx)/r$, $\omega_i\equiv
\kappa^0_\beta =\kappa^0_\alpha $ and $k_\omega\equiv \sqrt{\omega^2_i-m^2_i}$.
The crucial point here is that, depending on the masses and momenta of
the incoming particles, both neutrinos ($u\bar{u}$) and antineutrinos
($v\bar{v}$)  can contribute to the amplitude \eqref{A:4} depending on the sign
of its energy $\omega_i$, restricted to $|\omega_i|>m_i$; the off-shell
contributions for $\omega_i \in [-m_i,m_i]$ are exponentially decreasing and
then negligible for large distances (see appendix\,\ref{int}). We will see in
the following that antineutrino contributions in this case is possible  and it
corresponds to unphysical contributions.

We are interested in large production--detection separations. It permits us to
approximate, as in Ref.\,\cite{Dolgov}, $r\approx
R+\hat{\bR}.(\by-\bx_B)-\hat{\bR}.(\bx-\bx_A)$ and $\hat{\br}\approx \hat{\bR}$,
where $R\equiv |\bx_B-\bx_A|$ and $\hat{\bR}\equiv (\bx_B-\bx_A)/R$.
Such approximations inserted in Eq.\,\eqref{[]} lead to momentum conservation
on $x_A$ and $x_B$ vertices:
\begin{align}
2\pi\delta(\kappa^0_\beta -\kappa^0_\alpha )&
\frac{-i}{4\pi R}e^{ik_\omega R}
e^{-ik_\omega\hat{\bR}.(\bx_B-\bx_A)}
(2\pi)^3\delta^3(\bs{\kappa}_\beta -k_\omega\hat{\bR})
(2\pi)^3\delta^3(\bs{\kappa}_\alpha -k_\omega\hat{\bR})
\cr&\times
[u_i(k_\omega\hat{\bR})\bar{u}_i(k_\omega\hat{\bR})\theta(\omega_i-m_i)
-v_i(-k_\omega\hat{\bR})\bar{v}_i(-k_\omega\hat{\bR})\theta(-\omega_i-m_i)
]
\label{[]2}
~.
\end{align}

At this point we have all the information to analyze whether the antineutrino
part of the propagator contributes to the overall process.
Neither of the isolated processes $A+l_\alpha\rightarrow A'+\bar{\nu}_i$ and
$B+\bar{\nu}_i\rightarrow B'+l_\beta$ are allowed if we calculate the transition
amplitude for them separately using the weak Lagrangian \eqref{LW}.
(For Majorana neutrinos they are strongly suppressed by helicity mismatch.)
So far four-momentum conservation in both
$x_A$ and $x_B$ vertices were automatically required from the calculations;
among them the requirement of energy conservation for intermediate neutrinos
with respect to the accompanying particles in vertex $x_A$
($\omega_i=\kappa^0_\alpha $) and in vertex $x_B$ ($\omega_i=\kappa^0_\beta $),
is already implicit. The remaining are explicit in the delta functions of
Eq.\,\eqref{[]2}. The on-shell condition for neutrinos
($|\omega_i|^2-k_\omega^2=m^2_i$) for long distance propagation was also
automatic. What the calculations did not required is a definite sign for
$\omega_i$, for all possible momenta constrained by the mentioned
energy-momentum conservations. To analyze if and under what conditions both
signs are possible is equivalent to study the kinematics of two-body to two-body
scattering allowing the sign of one particle energy to be free. Putting in
equations, for vertex $x_A$, assuming the particle $A$ at rest, we obtain from
$(p_A-p_i)^2=(p_{A'}-p_\alpha )^2$ the neutrino energy \eq{ \label{Ei} E_i=
\frac{1}{2M_A}[M^2_A-M^2_{A'}+m^2_i-m^2_\alpha +2E_\alpha E_{A'}-2\bp_\alpha
.\bp'_{A}] ~. } The minimum value of right-hand side of Eq.\,\eqref{Ei}
corresponds to last two terms equal to $2m_\alpha M_{A'}$, which gives for the
minimum \eq{ \label{Ei:min}
\min(E_i)=
\frac{1}{2M_A}[M^2_A-(M_{A'}-m_\alpha )^2+m^2_i]
~.
}
The values $\omega_i=\kappa^0_\alpha$ are bounded from below by
the value in Eq.\,\eqref{Ei:min}. Imposing $\min(E_i)>m_i$ and $\min(E_i)<-m_i$
is respectively equivalent to
\eqarr{
\label{condM1}
M_A-M_{A'}&>&-(m_\alpha -m_i)\\
\label{condM2}
M_A-M_{A'}&<&-(m_\alpha +m_i)
~,
}
for $M_A>m_i$ and $M_{A'}>m_\alpha $. It is clear that depending on the value of
the masses, condition \eqref{condM2} may be satisfied leading to antineutrino
contributions to Eq.\,\eqref{A:4} for a range of possible incoming momenta.
Of course the condition \eqref{condM1} is sufficient to exclude antineutrino
contributions but it also excludes the cases where a threshold energy is
required for the lepton $l_\alpha$ to initiate the production reaction.
Thus to prevent antineutrino contributions, it is better to adopt the weaker
condition of restricting the sign of the energy of intermediate neutrinos
$\omega_i$ to be positive, keeping only the first term in Eq.\,\eqref{[]}.
Analogous analysis lead to possible momenta and mass values that allow
$\kappa^0_\beta<-m_i$ for vertex $x_B$, still compatible with
$\kappa^0_\alpha=\kappa^0_\beta$. Notice that condition $\kappa^0_\alpha>m_i$ is
exactly the kinematical condition to allow the production of physical neutrinos
in $x_A$ and $\kappa^0_\beta>m_i$ allow only the contribution of neutrinos with
energy above threshold to trigger the detection reaction.
The violation of these conditions implies in kinematically impossible
contributions in production or detection.

Restricted to condition $\omega_i>0$ we can insert the expression above into
Eq.\,\eqref{A:4} which yields
\eqarr{
\mcal{A}_i&=&
\int \widehat{d\bq_\alpha }
2\pi\delta(\kappa^0_\beta -\kappa^0_\alpha )\theta(\omega_i-m_i)
\frac{-i}{4\pi R}e^{ik_\omega R-i\omega_i(t_B-t_A)}
e^{i(p_k+p_B').x_B}e^{ip'_A.x_A}
\cr&&\times
u_k(\bp_\beta )\gamma^\mu L
u_i(k_\omega\hat{\bR})\bar{u}_i(k_\omega\hat{\bR})
\gamma^\nu L \psi_\alpha (\bq_\alpha )
\cr&&\times
\left.
J^{BB'}_\mu(\bq_B,\bp'_B)\frac{\psi_B(\bq_B)}{\sqrt{E_B(\bq_B)}}
J^{AA'}_\nu(\bq_A,\bp'_A)\frac{\psi_A(\bq_A)}{\sqrt{E_A(\bq_A)}}
\right|_{\parbox{18ex}{
$\mss{\bq_B=\bp_\beta +\bp'_B-k_\omega\hat{\bR}}$
\\[-.3\baselineskip]
$\mss{\bq_A=\bp'_A-\bq_\alpha +k_\omega\hat{\bR}}$
}}
\label{A:5}
\quad.
}
Notice that the step function $\theta(\omega_i-m_i)$ prevents non-physical
neutrinos to contribute to the process.

Particularly, if we use a unidimensional wave packet for the incoming
lepton $l_\alpha $
\eq{
\label{psi:j:delta}
\psi_\alpha (\bq)=\psi_\alpha (q_x,q_y,q_z)
=\delta(q_x)\delta(q_y)\psi_{\alpha z}(q_z)
~,
}
we obtain an amplitude analogous to Ref.\,\cite{Dolgov}
\eqarr{
\sum_iU_{\beta i}U^*_{\alpha i}\mcal{A}_i&=&\sum_i
\frac{-i}{4\pi R}e^{ik_\omega R-i\omega_i(t_B-t_A)}
2\pi
\left|2\mathrm{p}_\alpha \frac{\partial}{\partial\bq_\alpha ^2}(\kappa^0_\beta -\kappa^0_\alpha )
\right|^{-1}_{\bq_\alpha =\mathrm{p}_\alpha \hat{z}}
e^{i(p_k+p_B').x_B}e^{ip'_A.x_A}
\cr&&\times
U_{\beta i}u_\beta(\bp_\beta )\gamma^\mu L\,u_i(k_\omega\hat{\bR})
\,
U^*_{\alpha i}\bar{u}_i(k_\omega\hat{\bR})
\gamma^\nu L \psi_{\alpha z}(\mathrm{p}_\alpha )
\cr&&\times
\left.
J^{BB'}_\mu(\bq_B,\bp'_B)\frac{\psi_B(\bq_B)}{\sqrt{E_B(\bq_B)}}
J^{AA'}_\nu(\bq_A,\bp'_A)\frac{\psi_A(\bq_A)}{\sqrt{E_A(\bq_A)}}
\right|_{\parbox{19ex}{
$\mss{\bq_B=\bp_\beta +\bp'_B-k_\omega\hat{\bR}}$
\\[-.3\baselineskip]
$\mss{\bq_A=\bp'_A-\mathrm{p}_\alpha \hat{z}+k_\omega\hat{\bR}}$
}}
\label{A:6}
~,~~
}
where $\mathrm{p}_\alpha $ is the root of $f(|\bq_\alpha |=\mathrm{p}_\alpha )=\kappa^0_\beta -\kappa^0_\alpha =0$, which comes
from energy conservation from the whole process; if there is no root the
process is kinematically forbidden. The detection probability is proportional to
the square of the amplitude \eqref{A:6} integrated over the final phase space
$d\bp'_Ad\bp'_Bd\bp_\beta [2E_{A'}(\bp'_A)2E_{B'}(\bp'_B) 2E_\beta (\bp_\beta
)]^{-1}$. In particular, since $\bp_\beta ,\bp'_A,\bp'_B$ are fixed, the phases
that differ for different intermediate neutrinos $\nu_i$ are only $k_\omega
R-\omega_i(t_A-t_B)$ which is the same result obtained in
Ref.\,\onlinecite{Dolgov} (except for terms which depend on the mean velocity of
particles $A$ and $B$).

So far we have shown in an EWP approach both processes in $x_A$ and $x_B$ should
be considered real scattering processes with real neutrinos involved.
The off-shell contributions are negligible to large distances and antineutrino
contributions were explicitly excluded by eliminating the second
term of Eq.\,\eqref{[]}. These informations permit us to rewrite
Eq.\,\eqref{A:5} in a slightly different form
\eqarr{
-G^2
\sum_iU_{\beta i}U^*_{\alpha i}\mcal{A}_i&=&
\sum_i\int \frac{d\bp}{2E_i(\bp)}
\int d^4y\, \bra{B'(\bp'_B),l_\beta (\bp_\beta )}\lag_1(y)
e^{i(P-p_i).x_B}\ket{B,\nu_i(\bp)}
\cr&&\hs{6em}
\int d^4x\,\theta(y-x)
\bra{A'(\bp'_A),\nu_i(\bp)}\lag_1^\dag(x)
e^{iP.x_A}\ket{A,l_\alpha }
\label{A:7}
~,~~
}
where $P=(H,\mathbf{P})$ is the energy-momentum operator.
A change of notation were made here: in Eq.\,\eqref{A:6} the states
$\ket{B}$ and $\ket{A,l_\alpha }$ are centered around the origin while in
Eqs.\,\eqref{A:1}-\eqref{psi:j} they are respectively centered around $x_B$ and
$x_A$; the translation is explicitly performed by the translation operator
$e^{iP.x}$. Additionally, the step function $\theta(y-x)$ is necessary to ensure
that the contributions of points $y$ around $x_B$ should always be after
the contributions of points $x$ around $x_A$. By following the same steps from
Eq.\,\eqref{A:3} to Eq.\,\eqref{psi:j:delta} we can arrive from
Eq.\,\eqref{A:7} to Eq.\,\eqref{A:6}.

Equation \eqref{A:7} shows us the amplitude of the entire process from
production to detection in ``decomposed'' form (apart from the step
function in time): the amplitude of production process multiplied by the
amplitude of detection process summed over all possible intermediate real
neutrinos of different masses $m_i$ and momentum $\bp$. (The sum over spins are
implicit.)

\subsection{A simple second quantized formulation}
\label{sec:qft2}

Considering that only real neutrinos or antineutrinos (one of them
exclusively) should travel from production to detection, the possibility to use
the free second quantized theory for spin 1/2 fermions to describe flavor
oscillations is investigated in this section.
This simple, type B and QFT description of flavor oscillation phenomena
guarantees only particle or antiparticle propagation, keeping the nice property
of giving normalized oscillation probabilities, like the first quantized
examples treated in secs. \ref{sec:D} and \ref{sec:ST}.

To accomplish the task of calculating oscillation probabilities in QFT we have
to define the neutrino states that are produced and detected through weak
interactions.
Firstly, we define the shorthand for the combination of fields appearing in the
weak effective charged-current lagrangian \eqref{LW}
\eq{
\label{nu:f:field}
\nu_\alpha(x)\equiv
U_{\alpha i}\nu_i(x)~,~\alpha=e,\mu\,.
}
We will restrict the problem to two flavor families and use the matrix $U$ as
the same in Eq.~\eqref{mixing}.
The mass eigenfields $\nu_i(x)$, $i=1,2$, are the physical fields for which the
mass eigenstates $\ket{\nu_i(\bp)}$ are well defined asymptotic states.
The free fields $\nu_i(x)$ can be expanded in terms of creation and annihilation
operators (see appendix \ref{app:def}) and the projection to the one-particle
space defines the mass wave function \eq{ \label{wf:12}
\psi_{\nu_i}(\bx;g_i)=\bra{0}\nu_i(\bx)\ket{\nu_i\!:\!g_i}
\equiv
\underset{s}{\textstyle \sum}\int\!
d\bp\,\frac{g_i^s(\bp)}{\sqrt{2E_i}}u_i^s(\bx;\bp)
~,~~i=1,2,
}
where
\eq{
\ket{\nu_i\!:\!g_i}
\equiv
\underset{s}{\textstyle \sum}\int\!
d\bp\,g_i^s(\bp)\ket{\nu_i(\bp,s)}
~.
}

Since the creation operators for neutrinos (antineutrinos) can be written in
terms of the free fields $\bar{\nu}_i(x)$ ($\nu_i(x)$), we can define the
``flavor'' states as the superpositions of mass eigenstates
\eqarr{
\label{nu:f:qft}
\ket{\nu_\alpha\!:\!\{g\}}
&\equiv&
U^*_{\alpha i}
\ket{\nu_i\!:\!g_i}
\cr
\ket{\bar{\nu}_\alpha\!:\!\{g\}}
&\equiv&
U_{\alpha i}
\ket{\bar{\nu}_i\!:\!g_i}
~.
}
The details of creation are encoded in the functions $g_i$.

We can also define
\eqarr{
\label{wf:em}
\psi_{\nu_\alpha\nu_e}(x;\{g\})
&\equiv &
\bra{0}\nu_e(x)\ket{\nu_\alpha\!:\!\{g\}}
\cr
&=&
U_{ei}U^*_{\alpha i}\psi_{\nu_i}(x;g_i)
~,
}
where $\psi_{\nu_i}(x)$ are then mass wave functions defined in
Eq.~\eqref{wf:12}. We can see from Eq.~\eqref{wf:em} that if
$\psi_{\nu_1}(\bx,t)=\psi_{\nu_2}(\bx,t)=\psi(\bx)$, for a given time $t$,
$\psi_{\nu_e\nu_e}(\bx,t)=\psi(\bx)$ and $\psi_{\nu_\mu\nu_e}(\bx,t)=0$ due to
the unitarity of the mixing matrix.

Although this approach does not rely on flavor Fock spaces and Bogoliubov
transformations, we can use the same observables used by Blasone and Vitiello
to quantify flavor oscillation~\cite{BV:obs}: the flavor charges, which are
defined as
\eq{
\label{Q:f}
Q_\alpha(t)=\int\! d\bx:\nu_\alpha^\dag(\bx,t)\nu_\alpha(\bx,t):
~,~~\alpha=e,\mu\,,
}
where $:\;:$ denotes normal ordering.
Note that the $Q_e(t)+Q_\mu(t)=Q$ is conserved~\cite{BV:AP95},
the two flavor charges are compatible for equal times, i.e.,
$[Q_e(t),Q_\mu(t)]=0$, and
$\bra{\nu\!:\!\{g\}}Q\ket{\nu\!:\nolinebreak\!\{g\}}=\pm\braket{\nu\!:\!\{g\}}
{\nu\!:\!\{g\}}$ for any particle state (+) or antiparticle state (-). Notice
that in the second quantized version the charges can acquire negative values,
despite the fermion probability density in first quantization is a positive
definite quantity. The conservation of total charge guarantees the conservation
of total probability \eqref{prob:cons}.

We can further split the flavor charges into left-handed (-) and right-handed
(+) parts
\eq{
\label{Q:f:LR}
Q_\alpha^{(\pm)}(t)=\int\! d\bx
:\nu_\alpha^\dag(\bx,t) \mn{\frac{1}{2}}(\id\pm\gamma_5)\nu_\alpha(\bx,t):
~,~~\alpha=e,\mu\,,
}
where $Q_\alpha^{(+)}+Q_\alpha^{(-)}=Q_\alpha$. These components will be used to
calculate the left-handed to right-handed transition.

With the flavor charges defined, we can calculate, for example, the
conversion probability
\eqarr{
\label{prob:em:qft}
\mathscr{P}(\ms{\nu_e\!\rightarrow\!\nu_\mu};t)
&\equiv &
\bra{\nu_e\!:\!\{g\}}Q_\mu(t)\ket{\nu_e\!:\!\{g\}}
\\
&=&
U_{\mu i}U^*_{\mu j}U_{e j}U^*_{e i}
\int\! d\bp\,
e^{-i(E_i-E_j)t}
\tilde{\psi}_{\nu_j}^\dag(\bp;g_j)
\tilde{\psi}_{\nu_i}(\bp;g_i)
\label{prob:em:qft1}
~,
}
where the neutrino wave functions $\psi_{\nu_i}$ are defined in terms of the
function $g_i(\bp)$ in Eq.~\eqref{wf:12}.
If we could equate the two mass wavefunctions in momentum space
$\tilde{\psi}_{\nu_1}(\bp;g_1)= \tilde{\psi}_{\nu_2}(\bp;g_2)=
\tilde{\psi}_{\nu_e}(\bp)$ we would obtain, from Eq.\;\eqref{prob:em:qft1}, the
standard two family conversion probability \eqref{prob:em:S}
\eq{
\label{prob:em:qft2}
\mathscr{P}(\ms{\nu_e\!\rightarrow\!\nu_\mu};t)=
\int\!
d\bp\,\mathscr{P}(\bp,t)\tilde{\psi}_{\nu_e}^\dag(\bp)
\tilde{\psi}_{\nu_e}(\bp)
~,
}
where $\mathscr{P}$ was defined in Eq.~\eqref{P}.
However, the equality can not hold as proved in appendix~\eqref{app:dec}: two
wavefunctions with only positive energy components with respect to two bases
characterized by different masses can not be equal. Then, it is not
possible to impose a flavor definite condition.
Instead, we can write
\eq{
\label{g:psi}
g_i(\bp,s)=\frac{u_i^{s\dag}(\bp)}{\sqrt{2E_i(\bp)}}\tilde{\psi}_i(\bp)~,
}
where $\tilde{\psi_i(\bp)}$ is the initial wave function associated to the
neutrino of mass $m_i$ at creation, taking care to maintain the normalization
$\int d\bp\,|g_i(\bp)|^2=1$; any transition amplitude can be written in the form
Eq.\;\eqref{g:psi}. In general $\tilde{\psi}_i(\bp)=\tilde{\psi}(\bp,m_i)$, and
then, for small mass differences,
\eq{
\label{psi:1+d}
\tilde{\psi}_i(\bp)\approx \tilde{\psi}(\bp,\bar{m})\pm
\frac{\Delta m}{2}\frac{\partial}{\partial \bar{m}}\tilde{\psi}(\bp,\bar{m})
~,}
where $\bar{m}=(m_1+m_2)/2$ and $\Delta m=m_2-m_1$.
Keeping only the first term,
$\tilde{\psi}(\bp,\bar{m})\equiv\tilde{\psi}(\bp)$, we obtain from
Eq.\;\eqref{prob:em:qft1},
\eqarr{
\label{prob:em:qft3}
\mathscr{P}(\ms{\nu_e\!\rightarrow\!\nu_\mu};t)&=&
\int\!
d\bp\,\mathscr{P}(\bp,t)\tilde{\psi}^\dag(\bp)
[\id-\mn{\frac{1}{2}}\Lambda_{1-}(\bp)-\mn{\frac{1}{2}}\Lambda_{2-}(\bp)]
\tilde{\psi}(\bp)
\cr
&&+
\mn{\frac{1}{4}}\sin^22\theta
\int\!
d\bp\,\tilde{\psi}^\dag(\bp)
\big[f(\bp)\cos(\Delta Et)-i\frac{\Delta m}{2E_1E_2}\bs{\gamma}.\bp \sin(\Delta
Et) \big] \tilde{\psi}(\bp)
~.
}
Notice that in this case, the conversion probability is non-null for $t=0$,
\eq{
\mathscr{P}(\ms{\nu_e\!\rightarrow\!\nu_\mu};0)=
\mn{\frac{1}{4}}\sin^22\theta
\int\!
d\bp\,f(\bp)\tilde{\psi}^\dag(\bp)\tilde{\psi}(\bp)
~,
}
which imply a direct lepton flavor violation in creation. However, since
$f(\bp)\approx (\Delta m)^2/(4\bp^2)$ for ultra-relativistic momenta, the
violation is hopelessly feeble for direct measurement.
Among the deviations of the conversion probability \eqref{prob:em:qft3}
compared to the standard one \eqref{prob:em:qft2}, only the last term is of
order $\Delta m/\bar{E}$, the rest is of order $(\Delta m/\bar{E})^2$ (the
contributions of $\Lambda_{i-}$ can be estimated by
$[v_2^\dag(-\bp,s)u_1(\bp,s')]^2\sim \bp^2[\Delta m+\Delta
E]^2/[(m_1+E_1)(m_1+E_1)]$).
Even so, $\Delta m/\bar{E}\sim 10^{-9}$ for $\Delta m^2\sim
10^{-3}\mathrm{eV}^2$, $\bar{m}\sim 1/2{\rm eV}$ and $\bar{E}\sim
1\mathrm{MeV}$, which can not be seen in actual oscillation experiments.
Nevertheless, it is important to note that the knowledge of $\Delta m$ in
conjunction with $\Delta m^2$ gives information about the absolute mass scale
because of $\Delta m^2=2\bar{m}\Delta m$.

Using $Q_\alpha^{(-)}$ of Eq.\;\eqref{Q:f:LR} instead of $Q_\alpha$ in
Eq.\;\eqref{prob:em:qft1} and $\tilde{\psi}(\bp)=L\tilde{\psi}(\bp)$ in
Eq.\;\eqref{prob:em:qft2} we obtain
\eq{
\label{prob:em:qft:LR}
\mathscr{P}(\ms{\nu_{eL}\!\rightarrow\!\nu_{\mu R}};t)=
\int\!
d\bp\,\bigg[
\frac{m_1m_2}{4E_1E_2}\mathscr{P}(\bp,t)
+\mn{\frac{1}{4}}\sin^22\theta\Big(\frac{m_1}{2E_1}-\frac{m_2}{2E_2}\Big)^2
\bigg]\tilde{\psi}^\dag(\bp)\tilde{\psi}(\bp)
~.
}
The total probability loss from the conversion of initial left-handed electron
neutrino to right-handed neutrinos yields
\eq{
\label{prob:qft:LR}
\mathscr{P}(\ms{\nu_{eL}\!\rightarrow\!\nu_{eR}};t)+
\mathscr{P}(\ms{\nu_{eL}\!\rightarrow\!\nu_{\mu R}};t)=
\int d\bp\,\Big[\cos^2\!\theta\Big(\frac{m_1}{2E_1}\Big)^2
+\sin^2\!\theta\Big(\frac{m_2}{2E_2}\Big)^2\Big]
\tilde{\psi}^\dag(\bp)\tilde{\psi}(\bp)
~.
}
Notice Eq.\;\eqref{prob:qft:LR} does not depend on time in contrast to its
first quantized analog in Eq.\;\eqref{KD:LR}.
Other conversion and survival probabilities can be obtained from
Eq.\;\eqref{prob:cons:D} and
\eq{
\mathscr{P}(\ms{\nu_{eL}\!\rightarrow\!\nu_{\mu R}};t)+
\mathscr{P}(\ms{\nu_{eL}\!\rightarrow\!\nu_{\mu L}};t)
=\mathscr{P}(\ms{\nu_e}\!\rightarrow\!\nu_{\mu};t)
~.
}
The exchange of $L\leftrightarrow R$ does not modify the formulas, provided we
also change the chirality of the initial wave function.

For completeness we calculate the additional conversion probabilities including
the second term of Eq.\;\eqref{psi:1+d}
\eqarr{
\label{prob:em:qft:1+d}
\delta\mathscr{P}(\ms{\nu_{e}\!\rightarrow\!\nu_{\mu }};t)&=&
\mn{\frac{1}{4}}\sin^22\theta\frac{\Delta m}{2}
\int\!
d\bp\,\tilde{\psi}^\dag(\bp)\bigg[
\frac{H_2}{2E_2}-\frac{H_1}{2E_1}
+\frac{\Delta m}{2E_1E_2}\bs{\gamma}.\bp
\cr&&
+(\Lambda_{1+}+\Lambda_{2+}-f(\bp))i\sin\Delta Et
\bigg]\frac{\partial}{\partial \bar{m}}\tilde{\psi}(\bp)
+h.c.
~,
\\
\label{prob:em:qft:1+d:LR}
\delta\mathscr{P}(\ms{\nu_{eL}\!\rightarrow\!\nu_{\mu R}};t)&=&
\mn{\frac{1}{4}}\sin^22\theta\frac{\Delta m}{2}
\int\!
d\bp\,\tilde{\psi}^\dag(\bp)\bigg[
\big(\frac{m_2}{2E_2}\big)^2-\big(\frac{m_1}{2E_1}\big)^2
\cr&&~
+\frac{m_1m_2}{2E_1E_2}i\sin\Delta Et
\bigg]\frac{\partial}{\partial \bar{m}}\tilde{\psi}(\bp)
+h.c.~,
}
which have terms of order $\Delta m$ and $(\Delta m)^2$.

To calculate the conversion probability for antineutrinos
$\bar{\nu}_e\rightarrow\bar{\nu}_\mu$, it is enough to use
\eq{
\label{g*:psi}
g_i^{s*}(\bp)\equiv
\tilde{\psi}_i^{\dag}(\bp)\frac{v_i^s(\bp)}{\sqrt{2E_i(\bp)}} ~,
}
instead of Eq.\;\eqref{g:psi}, replace $t\rightarrow -t$ and
$\tilde{\psi}(\bp)\rightarrow \frac{\bs{\alpha}.\bp}{|\bp|}\tilde{\psi}(\bp)$
in the expressions corresponding to neutrinos
\eqref{prob:em:qft1}--\eqref{prob:em:qft:1+d:LR}, or apply charge conjugation
$\tilde{\psi}(\bp)\rightarrow -\gamma_0C\tilde{\psi}^*(\bp)$. These
prescriptions can be inferred from direct comparison to the calculations and
beware that the definition of antineutrino states are defined with
$g_i^{s*}(\bp)$ \eqref{ket:nub:t}.

The formulas obtained in this second quantized version does not have the
interference terms between positive and negative energies like in
Eq.\;\eqref{prob:em:D2}.  Such interference terms are absent because the
possible mixed terms like $b_2(\bp)a^\dag_1(\bp)\ket{0}$ are null for an initial
``flavor'' state superposition that contains only particle states (or only
antiparticles states). Notice that the irrelevance of the initial spinorial
structure no longer holds in this second quantized version,
which can be seen, for example, in Eq.\;\eqref{prob:em:qft3}.

\section{Discussion and Conclusions}

Using the Dirac equation which is a relativistic covariant equation we obtained
oscillation probabilities respecting causal propagation.
The oscillation formulas obtained had additional rapid oscillating terms
depending on the frequency $E_1+E_2$, with respect to the usual oscillation
formulas with wave packets.
Such additional oscillatorial character seemed to have its origin
in the intrinsic spinorial character used. However, we have seen that such terms
also appear in the charged spin 0 particle oscillations.
In fact the rapid oscillatorial terms arise from the interference of positive
and negative frequency parts of the initial WP and they are always present
independently of the initial wave packet if initially a flavor definite
condition were imposed.
In addition, within Dirac theory, we have shown the detailed spinorial character of the
initial wave function was irrelevant for flavor oscillation, independently of
the momenta involved.
The inclusion of the left-handed nature of
the created and detected neutrinos could also be simply achieved.
It is important to stress that the modifications found in this context would
have tiny observable effects to the flavor oscillation of ultrarelativistic
neutrinos.

Regarding second quantized approaches (sec.\;\ref{sec:qft}), in particular, EWP
approaches, we can compare the propagators used in the latter to the
evolution kernels in IWP approaches.
Both the free evolution kernel  and the Feynman propagator for fermions contain
the contribution from particle and antiparticles. From this perspective,  EWP
approaches could also contain both contributions from neutrinos and
antineutrinos, as in the first quantized approaches presented in
secs.\;\ref{sec:D} and \ref{sec:ST}.
To analyze this point, an EWP calculation
were carried out explicitly in sec.\,\ref{sec:qft} following
Ref.\,\onlinecite{Dolgov}.
Restricted to a case where only neutrinos would be present,
the calculation showed that the antineutrino contribution were not excluded
automatically in the formalism but a subsidiary condition could be necessary:
the sign of the frequency of the intermediate neutrinos should be restricted to
be positive.
In such case, there can be interference terms between positive
and negative frequencies, possibly yielding rapid oscillation terms similar to
the ones obtained in Dirac theory of sec.\;\ref{sec:D}. However, it should be
stressed that the origin of the interference between positive and negative terms
are different in the first quantized Dirac theory treated in sec.\;\ref{sec:D}
and in the EWP (second quantized) treated in this sec.\;\ref{sec:qft}. The
former comes from causality and completeness arguments in a classical field
theory perspective, while the latter have its origin in the consideration of 
non-physical contributions.
The restriction implied by the subsidiary condition automatically guarantee
that: (i) only real neutrinos that are kinematically allowed in production
contributes and (ii) in detection, only neutrinos with energies above threshold
to trigger the detection reaction contribute. Otherwise, kinematically forbidden
reactions in production or detection could be possible through exchange of
virtual antineutrinos carrying negative energy. The overall
energy-momentum conservation, though, is always respected (smeared out through
finite momentum distributions) through production/propagation/detection
processes. Since the presence of both neutrino--antineutrino contributions is
common to all EWP approaches, the subsidiary condition necessary in the EWP
approach analyzed is possibly necessary in any approach with virtual neutrino
propagation. (Unless a stronger condition like Eq.\;\eqref{condM1} is already
implicit.)
For example, in Eq.\;(14) of Ref.\;\onlinecite{GS}, the subsidiary condition
(for antineutrinos) is satisfied because the detection reaction is an elastic
scattering.  (Although the detection electrons are off-shell (bound state),
the subsidiary condition is valid.) An important remark in this respect is that
the calculations of the production and detection amplitudes as separate
processes take automatically into account the physical kinematical conditions
(i) and (ii) through the energy-momentum delta functions. It is also important
to stress that the result obtained here did not depend on particular wave
packets neither on the particular interaction used. The interesting point is
that by imposing such subsidiary condition beforehand enables us to write the
transition amplitude for the entire production/propagation/detection process in
a ``decomposed'' form, with simple physical interpretation. A more detailed
discussion about the decomposition of the process in separate production,
propagation and detection processes can be found in
Ref.\;\onlinecite{Cardall.00}. Similar conclusions can be drawn for the case
where only antineutrinos should propagate: the sign to be chosen should be
negative. A realistic EWP approach for antineutrino propagation is given in
Ref.\;\onlinecite{GS}. To conclude this part, EWP approaches for neutrino
oscillations require for consistency, but do not automatically imply, real
intermediate neutrinos or antineutrinos exclusively.

All the properties discussed above about EWP approaches suggest that the
description of two macroscopically distant scattering processes (production and
detection) as a single scattering process described by a single scattering
matrix have to be treated with care. We can be confident about the use of the
$S$ matrix to describe any microscopic event through perturbation theory to any
order of expansion (if the theory is renormalizable), but the extension to
macroscopically distant reactions is not automatic. Actually, if the two
processes are indeed not causally connected it can be proved that the $S$
matrix decomposes as the product of the two $S$ matrices for the two distant and
independent processes~\cite{Weinberg}.

From the considerations above, the positive and negative interference terms in
the first quantized approaches considered seem unphysical.
To support this idea, it was shown in sec.\,\ref{sec:qft2} that a simple second
quantized, type B, and IWP-like, approach could be devised using the second
quantized free theory  maintaining the simple properties of IWP approaches
but eliminating the undesirable interference of positive and negative
frequencies that was inevitable in the relativistic quantum mechanical context.
On the other hand, new ingredients such as the direct flavor violation in
creation and deviations from the standard oscillation formula were found.
The deviations include the probability loss due to the conversion of left-handed
neutrinos to right-handed neutrinos. Unfortunately, those new effects are tiny
either because of the small mass difference or the ultrarelativistic nature of
neutrinos and they are not feasible for direct observation in actual oscillation
experiments.

\acknowledgments
This work was supported by Conselho Nacional de Desenvolvimento
Cient\'\i fico e Tecnol\'ogico (CNPq).
The author would like to thank Prof.~J.~C.~Montero for a critical reading of the
manuscript.

\appendix
\setlength{\baselineskip}{.8\baselineskip}
\section{Notation and definitions}
\label{app:def}

The (scalar, spinorial or ST) wave functions
related by Fourier transforms are denoted as
\eqarr{
\label{fourier:xp}
\varphi(\bx)
&=&\frac{1}{(2\pi)^{\mt{3/2}}}
\int\! d\bp\,\tilde\varphi(\bp)\,e^{i\bp.\bx}
~,
\\
\label{fourier:px}
\tilde\varphi(\bp)
&=&\frac{1}{(2\pi)^{\mt{3/2}}}
\int\! d\bx\,\varphi(\bx)\,e^{-i\bp.\bx}
~.
}
The tilde denotes the inverse Fourier transformed function.

Using the property of the Dirac or ST Hamiltonian, $H_n^2=(\bp^2+m_n)^2\id$, we
can write the evolution operator in the form
\eq{
e^{-iH_nt}=\cos(E_nt)-i\frac{H_n}{E_n}\sin(E_nt)
~,
}
where the momentum dependence have to be replaced by $-i\nabla$ in coordinate
space.

The free neutrino field expansion used is ($i=1,2$)
\eq{
\label{psi:qft}
\nu_i(x)=
\underset{s}{\textstyle \sum}\int\! \frac{d\bp}{2E_{\bp}}\,
[u_i^s(x;\!\bp)a_i(\bp,s)+v_i^s(x;\!\bp)b_i^\dag(\bp,s)]
~,
}
where the creation and annihilation operators satisfy the
canonical anticommutation relations
\eqarr{
\{a_i(\bp,r),a_j^\dag(\bp',s)\}&=&\delta_{ij}\delta_{rs}2E_i(\bp)
\delta^3(\bp-\bp ')
~,
\\
\{b_i(\bp,r),b_j^\dag(\bp',s)\}&=&\delta_{ij}\delta_{rs}2E_i(\bp)
\delta^3(\bp-\bp ')~;
}
all other anticommutation relations are null.
The functions $u,v$ are defined as
\eqarr{
\label{u:x}
u_i^s(x;\!\bp)&=&u_i^s(\bp)\frac{e^{-ip_i.x}}{(2\pi)^{\mt{3/2}}}
~,
\\
\label{u:p}
u_i^s(\bp)&=&\frac{m_i+E_i\gamma^0-\bp.\bs{\gamma}}{\sqrt{E_i+m_i}}u_0^s
~,
\\
\label{v:x}
v_i^s(x;\!\bp)&=&v_i^s(\bp)\frac{e^{ip_i.x}}{(2\pi)^{\mt{3/2}}}
~,
\\
\label{v:p}
v_i^s(\bp)&=&\frac{m_i-E_i\gamma^0+\bp.\bs{\gamma}}{\sqrt{E_i+m_i}}v_0^s~,
}
where $p_i.x=E_i\ms{(\bp)}t-\bp.\bx$ and they satisfy the properties
\eqarr{
\bar{u}_0^ru_0^s=u_0^{r\dag}u_0^s&=&
-\bar{v}_0^rv_0^s=v_0^{r\dag}v_0^s=\delta_{rs}
~,
\\
v_0^{r\dag}u_0^s&=&u_0^{r\dag}v_0^s=0 ~~\forall~ r,s
~,
\\
\label{uv:complete}
\underset{s}{\textstyle \sum}u_i^{s}(\bp)\bar{u_i}^s(\bp)
&=&{\sla p + m_i}
=2E_i(\bp)\Lambda^D_{i+}(\bp)\gamma^0
~,
\\
\underset{s}{\textstyle \sum}v_i^{s}(\bp)\bar{v_i}^s(\bp)
&=&{\sla p - m_i}
=2E_i(\bp)\Lambda^D_{i-}(-\bp)\gamma^0
~.
}

The Feynman propagator for fermions is
\eqarr{
iS_F(x-y)&\equiv& \bra{0}T(\psi(x)\bar{\psi}(y))\ket{0}
\\
&=&\int\! \frac{d^4\!p}{\ms{(2\pi)}^4}\,\frac{i}
{\sla{p}-m+i\epsilon}\,e^{-ip.(x-y)}
\\
&=&(i\sla{\partial}+m)i\Delta_F(x-y;m)
~.
}

The function $S$ in Eq.~\eqref{S} and its equivalent for the Sakata-Taketani
Hamiltonian can be written as
\eqarr{
\label{S:D}
iS(x;m)&=&(i\sla\partial+m)i\Delta(x;m)
~,
\\
\label{S:ST}
K^{\ST}(x;m)&=&[i\partial_t-\frac{\nabla^2}{2m}(\tau_3\!+\!i\tau_2)+m^2]
i\Delta(x; m )
~,
\\
\label{Delta}
i\Delta(x;m)&=&\frac{1}{(2\pi)^3}\int d^4p\,
\delta(p^2-m^2)\epsilon(p_0)e^{-ip.x}
\cr
&=&
\frac{1}{(2\pi)^3}\int
\frac{d\bp}{2E_p}\,[e^{-ip.x}-e^{+ip.x}]
~.
}

The free neutrino eigenstates are defined as
\eqarr{
\ket{\nu_i(\bp,s)}&\equiv &\frac{a_i^\dag(\bp,s)}{\sqrt{2E_i}}\ket{0}
\\&=&
\int d\bx\,\nu_i^\dag(x)\ket{0}\frac{u_i(x;\bp)}{\sqrt{2E_i}}
~,
\\
\ket{\bar{\nu}_i(\bp,s)}&\equiv&\frac{b_i^\dag(\bp,s)}{\sqrt{2E_i}}\ket{0}
\\&=&
\int d\bx\,\frac{v_i^\dag(x;\bp)}{\sqrt{2E_i}}\nu_i(x)\ket{0}
~,
}
whose normalization is
$\braket{\nu_j(\bp,r)}{\nu_i(\bp',s)}=\delta_{ij}\delta_{rs}
\delta^3(\bp -\bp')$. The same normalization is valid for the antiparticle
states.
The states with finite momentum distributions are defined as
\eqarr{
\ket{\nu_i\!:\!g}&=&
\underset{s}{\textstyle \sum}\int\!
d\bp\,g^s(\bp)\ket{\nu_i(\bp,s)}
\\&=&
\int d\bx\,\nu_i^\dag(x)\ket{0}\psi_{\nu_i}(x)~,
\\
\label{psi:nu}
\psi_{\nu_i}(x)&\equiv&
\underset{s}{\textstyle \sum}\int\!
d\bp\,g^s(\bp)\frac{u_i^s(x;\bp)}{\sqrt{2E_i}}~,
\\
\label{ket:nu:t}
e^{-iHt}\ket{\nu_i\!:\!g}&=&
\int d\bx\,\nu_i^\dag(\bx,0)\ket{0}\psi_{\nu_i}(\bx,t)
~,\\
\label{ket:nub:t}
\ket{\bar{\nu}_i\!:\!g}&=&
\underset{s}{\textstyle \sum}\int\!
d\bp\,g^{s*}(\bp)\ket{\bar{\nu}_i(\bp,s)}
\\&=&
\int d\bx\,\psi_{\bar{\nu}_i}^\dag(x)\nu_i(x)\ket{0}~,
\\
\label{psi:nub}
\psi_{\bar{\nu}_i}(x)&\equiv&
\underset{s}{\textstyle \sum}\int\!
d\bp\,g^{s}(\bp)\frac{v_i^s(x;\bp)}{\sqrt{2E_i}}
~,
\\
e^{-iHt}\ket{\bar{\nu}_i\!:\!g}&=&
\int d\bx\,\psi_{\bar{\nu}_i}^\dag(\bx,t)\nu_i(\bx,0)\ket{0}
~.
}

\section{Decomposition with respect to two bases}
\label{app:dec}

It is possible to decompose a given spinorial function $\psi(\bx)$ in terms of
bases depending on different masses $m_1$ and $m_2$. Equating
\eq{
\label{psi:dec}
\psi(\bx)=
\int\! \frac{d\bp}{2E_i}\,[u_i^s(\bx;\bp)g_i^{\mt{(+)}}(\bp,s)
+v_i^s(\bx;\bp)g_i^{\mt{(-)}}(\bp,s)]
~,~~i=1,2\,,
}
the expansion coefficients can be obtained
\eqarr{
g_i^{\mt{(+)}}(\bp,s)&=&\int\! d\bx\,u_i^{s\dag}(\bx;\bp)\psi(\bx) ~,\\
\label{psi:coef}
g_i^{\mt{(-)}}(\bp,s)&=&\int\! d\bx\,v_i^{s\dag}(\bx;\bp)\psi(\bx)
~.
}
From Eq.~\eqref{psi:coef} we see that imposing the conditions
\eqarr{
g_1^{\mt{(-)}}(\bp,s)&=&
v_1^{s\dag}(\bp)\tilde{\psi}(-\bp)=0
~,
\\
g_2^{\mt{(-)}}(\bp,s)&=&
v_2^{s\dag}(\bp)\tilde{\psi}(-\bp)=0
~,
}
for all $\bp$, lead to the equivalent conditions
\eqarr{
[(m_1+E_2)-(m_2+E_2)]\,v_0^{s\dag}\tilde{\psi}(-\bp)&=&0
~,~s=1,2\,,
\label{cond1}
\\
\label{cond2}
\bigg[\frac{1}{m_1+E_2}-\frac{1}{m_2+E_2}\bigg]\,
v_0^{s\dag}\bs{\gamma}.\bp\tilde{\psi}(-\bp) & = & 0 ~,~s=1,2\,,
}
where the property of Eq.~\eqref{v:p} and $\gamma_0v_0^s=-v_0^s$ were
used.
In case $m_1\neq m_2$, we can use the decomposition
$\tilde{\psi}(\bp)=\tilde{\psi}_+(\bp)+\tilde{\psi}_-(\bp)$, where
$\tilde{\psi}_\pm(\bp)=(\id\pm\gamma_0)\tilde{\psi}(\bp)$/2,
and obtain from Eqs.~\eqref{cond1} and \eqref{cond2} the conditions
\eqarr{
v_0^{s\dag}\tilde{\psi}_-(-\bp)&=&0
~,~s=1,2\,,
\label{cond1:2}
\\
\label{cond2:2}
u_0^{s\dag}\bs{\sigma}.\bp\tilde{\psi}_+(-\bp) & = & 0 ~,~s=1,2\,,
}
where the properties $\bs{\gamma}=\gamma_0\gamma_5\bs{\sigma}$ and
$u_0^s=\gamma_5v_0^s$ were used in Eq.~\eqref{cond2:2}.
The equations \eqref{cond1:2} and \eqref{cond2:2} are only satisfied if
$\tilde{\psi}_+(\bp)=\tilde{\psi}_-(\bp)=0$, since $\bs{\sigma}.\bp$ has only
non-null eigenvalues and it commutes with $\gamma_0$. It is easier to reach this
conclusion in the helicity basis $\{u_0^{\mt{(\pm)}},v_0^{\mt{(\pm)}}\}$
characterized by $\bs{\sigma}.\bp u_0^{\mt{(\pm)}}=\pm |\bp| u_0^{\mt{(\pm)}}$,
but the result is basis independent.

\section{Integrals}
\label{int}

Splitting the Feynman propagator into positive and negative frequency parts
$iS_F(x)=iS^{\mt{(+)}}(x)+iS^{\mt{(-)}}(x)$ the following integrals give us
\eqarr{
\label{wS:+}
\int dt\,e^{i\omega t}iS^{\mt{(+)}}(\br,t;m)
&=&
(-i)\frac{e^{ik_\omega r}}{4\pi r}
[(\omega\gamma_0-k_\omega(\hat{\br}.\bs{\gamma)}+m]\theta(\omega -m)
\\
\label{wS:-}
\int dt\,e^{i\omega t}iS^{\mt{(-)}}(\br,t;m)
&=&
(-i)\frac{e^{ik_\omega r}}{4\pi r}
[(\omega\gamma_0-k_\omega(\hat{\br}.\bs{\gamma)}+m]\theta(-\omega -m)
~,
}
where $k_\omega=\sqrt{\omega^2-m^2}, \br=r\hat{\br}$, the conditions
$mr,k_\omega r\gg 1$ were assumed and terms behaving as $1/r^2$ were neglected.

To illustrate the calculations, Eq.\,\eqref{wS:+} is obtained by
\eqarr{
\int dt\,e^{i\omega t}iS^{\mt{(+)}}(\br,t;m)
&=&
\label{wS:1}
\frac{1}{(2\pi)^3}\int \frac{d\bp}{2E(\bp)}
\frac{i(E(\bp)\gamma_0-\bp.\bs{\gamma}+m)}
{\omega-E(\bp)+i\epsilon}e^{i\bp.\br}
\\
&=&
\frac{1}{(2\pi)^2}\frac{-i}{2r}
\int_{-\infty}^{\infty}
dp\frac{p}{E(p)}
\Big\{
\sin{pr}
\frac{E(p)\gamma_0+m}{E(p)-\omega-i\epsilon}
\cr&&
\hs{5em}+~
[\cos{pr}-\frac{\sin{pr}}{pr}]
\frac{ip(\hat{\br}.\bs{\gamma})}{E(p)-\omega-i\epsilon}
\Big\}
\label{wS:2}
~.
}
In Eq.\,\eqref{wS:1} the following identity is used
\eq{
\int_0^{\infty}dt\,e^{\pm iEt}=\frac{\pm i}{E\pm i\epsilon}
~.
}
To get to the result of Eq.\,\eqref{wS:+} it is necessary to split the
functions $\sin pr$ and $\cos pr$  in Eq.\,\eqref{wS:2} into exponentials
and, for the $e^{ipr}$ part, integrate along a closed path formed by a half
semicircle in the upper-half complex plane going round the branching
line $[im,i\infty)$ (for the $e^{-ipr}$ part take the path reflected by
the line defined by $\mathrm{Re}z=0$).
The contribution from the paths beside the branching line yields
a function which decreases more rapidly than $e^{-mr}$ and it is negligible for
$mr\gg 1$. The contributions for $-m<\omega<m$ give a function with
dependence $e^{-|k_\omega|r}$ which is also negligible for large separations
$r$.



\begin{thebibliography}{99}
\bibitem{endnote1}
The charge here is a global U(1) charge which is conserved for $n=1,2$ and does
not mean eletric charge. Obviously the ``flavor'' charges $\alpha=e,\mu$,
introduced by mixing, depend on time and quantifies the oscillation, as in
Eq.~\eqref{prob:em:ST}.
\end{thebibliography}
\end{document}